\documentclass[aps, showkeys]{revtex4}
\usepackage{}
\usepackage{graphicx}% Include figure files
\usepackage{dcolumn}% Align table columns on decimal point
\usepackage{bm}% bold math
\usepackage{epstopdf}
\usepackage{mathtools}
\usepackage{natbib}
\usepackage{captcont}

\begin{document}

\title{Limitation of the Least Square Method in the Evaluation of Dimension of Fractal Brownian Motions}
\author{
BINGQIANG QIAO$^{1\,,2}$, SIMING LIU$^{* \,2}$, HOUDUN ZENG$^2$, XIANG LI$^2$, and BENZHONG DAI$^1$
%\and Emma de O\~na Wilhelmi\inst{1}
%\and Felix Aharonian\inst{1, 3}
}
\affiliation{
$^1$ Department of Physics, Yunnan University, Kunming, China, 650091. \\
$^2$ Key Laboratory of Dark Matter and Space Astronomy, Purple Mountain Observatory, Chinese Academy of Sciences, Nanjing, 210008, China, liusm@pmo.ac.cn}

\begin{abstract}
{With the standard deviation for the logarithm of the re-scaled range $\langle |F(t+\tau)-F(t)|\rangle$ of simulated fractal Brownian motions $F(t)$ given in a previous paper \cite{q14}, the method of least squares is adopted to determine the slope, $S$, and intercept, $I$, of the log$(\langle |F(t+\tau)-F(t)|\rangle)$ vs $\rm{log}(\tau)$ plot to investigate the limitation of this procedure. It is found that the reduced $\chi^2$ of the fitting decreases with the increase of the Hurst index, $H$ (the expectation value of $S$), which may be attributed to the correlation among the re-scaled ranges. Similarly, it is found that the errors of the fitting parameters $S$ and $I$ are usually smaller than their corresponding standard deviations. These results show the limitation of using the simple least square method to determine the dimension of a fractal time series. Nevertheless, they may be used to reinterpret the fitting results of the least square method to determine the dimension of fractal Brownian motions more self-consistently. The currency exchange rate between Euro and Dollar is used as an example to demonstrate this procedure and a fractal dimension of 1.511 is obtained for spans greater than 30 transactions.

}
\end{abstract}


\maketitle

\section{Introduction}
The properties of fractional Brownian motions (fBms) have been investigated by researchers in different fields, e.g. statistics, hydrology, biology, finance, public transportation and so on, which has helped us to better understand many complex time series observed in nature \cite{m68,p78,l99, a02}.
%Fractals are a basic tool to describe the shape of the coastlines and the mountains and rivers, %the shape of a given time series is similar to that of these coastlines and mountains and %rivers.
The Hurst exponent $H$ ($0<H<1$) is the most important parameter characterizing any given time series $F(t)$, where $t$ represents the time steps, and the fractal dimension $D$ is
determined via the relation $D=2-H$.
%Useful model provided by fBms is a good tool to better understand many natural time series %\cite{q14}.
%According to Qiao, B. Q $\&$ Liu, S. M, (2013),
The Hurst exponent $H$ is defined with the following expression\cite{q14}:
\begin{equation}
\textrm{log}(\langle|F(t+\tau)-F(t)|\rangle) \propto H \textrm{log}(\tau),
\end{equation}
where $'\langle\rangle'$ represents averaging over $t$. Using the Lowen method for $0<H<0.5$ and the circulant embedding method for $0.5<H<1$, Qiao \& Liu (2013) carried out extensive simulations of fBms with different $H$ to obtain the standard deviation of the re-scaled range $\langle|\bigtriangleup F_\tau|\rangle =  \langle|F(t+\tau)-F(t)|\rangle$ for different sampling methods. This paper extends this study to investigate the limitation of applying the commonly used least square (LS) method to the log$(\langle |F(t+\tau)-F(t)|\rangle)$ vs $\rm{log}(\tau)$ plot for the evaluation of $H$.

This paper is organized as follows. With the sampling method 4 of Qiao \& Liu (2013), the standard LS method with weight is applied to the $\textrm{log}(\langle|F(t+\tau)-F(t)|\rangle)$ vs $\textrm{log}(\tau)$ plot for a few fBms with different $H$ in Section $2$, and the results of the fitting are analyzed. In Section $3$, a similar work is carried out for the third
sampling method of Qiao \& Liu (2013). Finally, we give our conclusion and a brief discussion in Section $4$.

\section{The LS fitting for one sampling method}

\begin{figure*}[htb]
\centering
\includegraphics[height=80mm,angle=0]{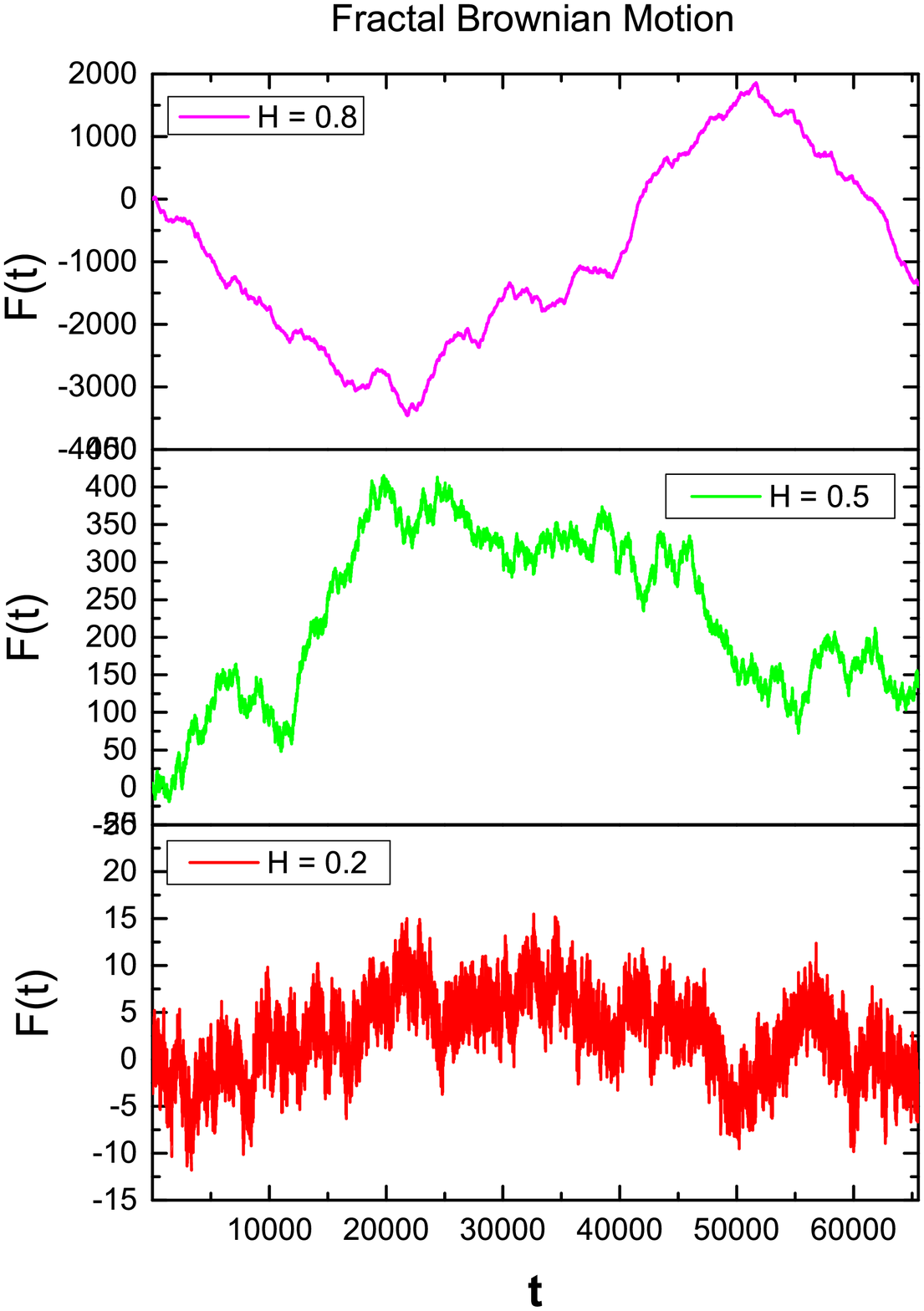}
\includegraphics[height=80mm,angle=0]{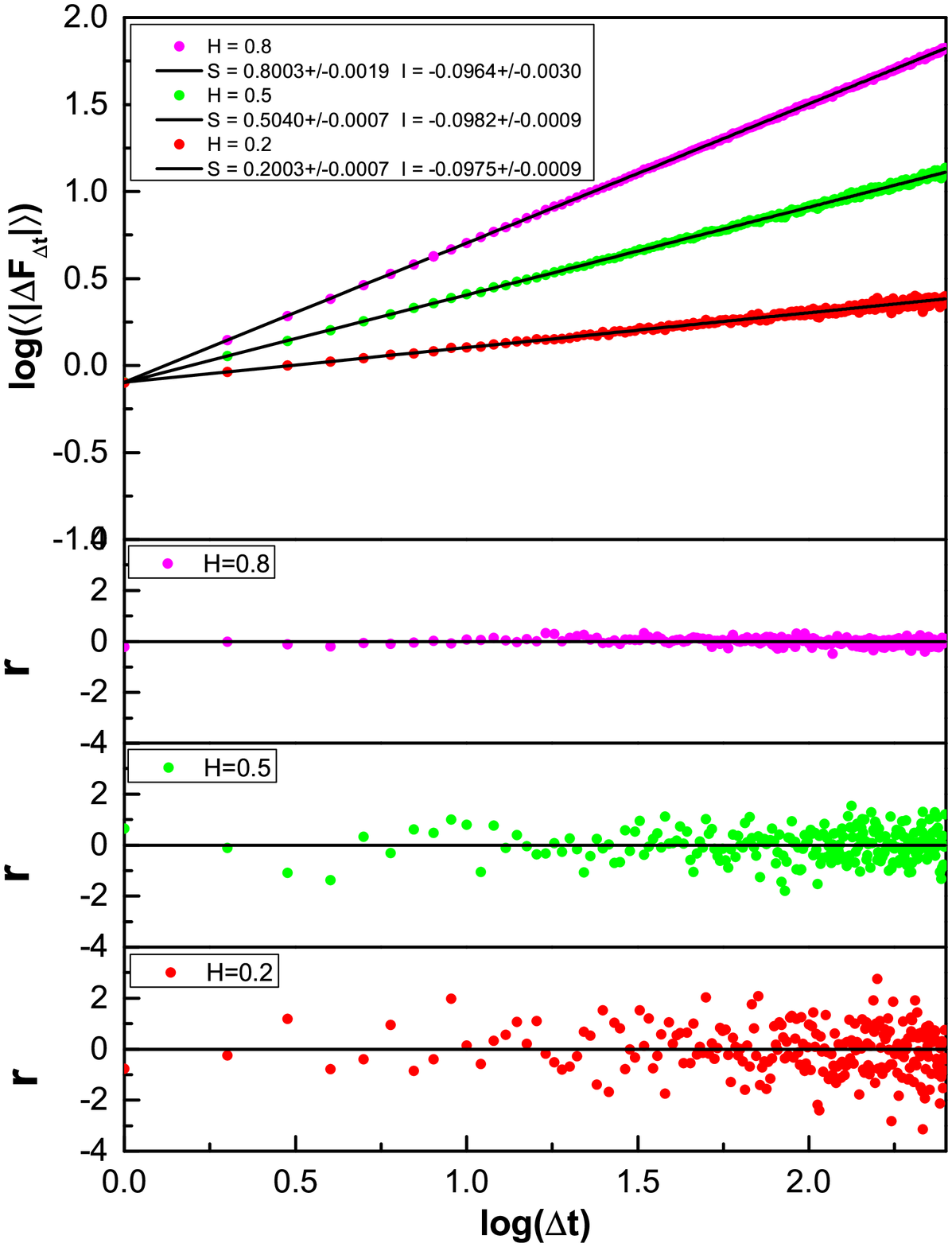}
\caption{Left: examples of a few fBms with different values of $H$. Right: dependence of the re-scaled range $\langle|\Delta F_{\Delta t}|\rangle$ on the time span $\Delta t$ for $H=0.8$ (magenta), $H=0.5$ (green), $H=0.2$ (red). The three black solid lines give the best fit obtained with the standard LS method. The fitting parameters and errors are indicated in the figure. The normalized residuals $r$ of $\textrm{log}(\langle|\Delta F_{\Delta t}|\rangle)$ for different $H$ values are shown in the three panels below.
 %are shown on the bottom figure (corresponding to the three situations on the top).
}
\label{a}
\end{figure*}

In the following, we use discrete-time fBms $F(t)$ to obtain the dependence of
the rescaled range $\langle |\Delta F_{\Delta t}|\rangle =\langle |F(t+\Delta t)-F(t)|\rangle$
on the time span $\Delta t$. The sampling method corresponds to the case 4 of Qiao, B.Q and Liu, S.M, (2013), which has the lowest standard deviation for the rescaled range and can be expressed as
%\begin{equation}
%\langle |\Delta F_t|\rangle = ~\left\{
%             \begin{tabular}{ll}
%             $\displaystyle\sum_{i=0}^{{\rm Int}(N^{2}/\Delta t)-1} |F(\Delta t + i*\Delta t)-F(i*\Delta t)|\over {\rm Int}(N^{2}/\Delta t)$  ~~~~~   & $0 < H < 0.5$ \\
%             $\displaystyle\sum_{i=0}^{{\rm Int}(N^{2}/\Delta t)^{\beta}-1} |F(\Delta t + i*\Delta t)-F(i*\Delta t)|\over {\rm Int}(N^{2}/\Delta t)^{\beta}$  ~~~~~   & $0.5 < H < 1$
%             \end{tabular}
%             \right. ,
%\end{equation}
\begin{equation}
\langle|\Delta F_{\Delta t}|\rangle={\displaystyle\sum_{i=0}^{{\rm Int}(N^2/\Delta t)-1} [|F(\Delta t+\Delta t*i) - F(\Delta t*i)|]\over {\rm Int}(N^2/\Delta t)}
\end{equation}
where $N=250$ and $\Delta t$ is a positive integer.
According to our previous work \cite{q14}, the standard deviation of $\textrm{log}(\langle|\Delta
F_{\Delta t}|\rangle)$ is given by: $$\sigma = N^{-1} \textrm{log}(e) (\pi/2-1)^{1/2}(\Delta t)^{1/2}
 {\rm\ \ \  for\ \ \  }0<H\leq 0.5$$ and $$\sigma = N^{-\beta} \textrm{log}(e) (\pi/2-1)^{1/2}(\Delta t)^{\beta/2}{\rm \ \ \ for \ \ \ } 0.5<H<1.0$$
 where $\beta = 2-2H+0.26^{1/2}-[(2H-1.5)^2+0.01]^{1/2}$.
 It should be noted that all of the measurement errors are only in the $\textrm{log}(\langle |\Delta F_{\Delta t}|\rangle)$ variable and the values of $\triangle t$ are all exact without errors. Then the standard LS method with weight (e.g. \citep{c08}; chapter 15 of \citet{p07})
  %in which the distances between the fitted line and the simulative data in the $y-$direction are %minimized and the weighting factors pertain to $y-$variables only, are
  can be applied to the $\textrm{log}(\langle|F(t+\Delta t)-F(t)|\rangle)$ vs $\textrm{log}(\Delta t)$ plot to obtain the slope $S$, the intercept $I$, and their corresponding errors $S_{e}$, $I_{e}$, respectively.
   %can be obtained, which represent the slope, the intercept, the slope's error and the %intercept's error of the fitting lines, respectively. The detail of fitting method can be found in the chapter 15 of \citet{p07}.
   Figure \ref{a} shows these results for several values of $H$.
   %best-fitting lines of the $\textrm{log}(\langle|F(t+\Delta t)-F(t)|\rangle)$ vs %$\textrm{log}(\Delta t)$ plot with different $H$, and the normalized residual $\chi$ of the %$\textrm{log}(\langle |\Delta F_t|\rangle)$ variable.

\begin{figure*}[htb]
\centering
\includegraphics[height=42mm,angle=0]{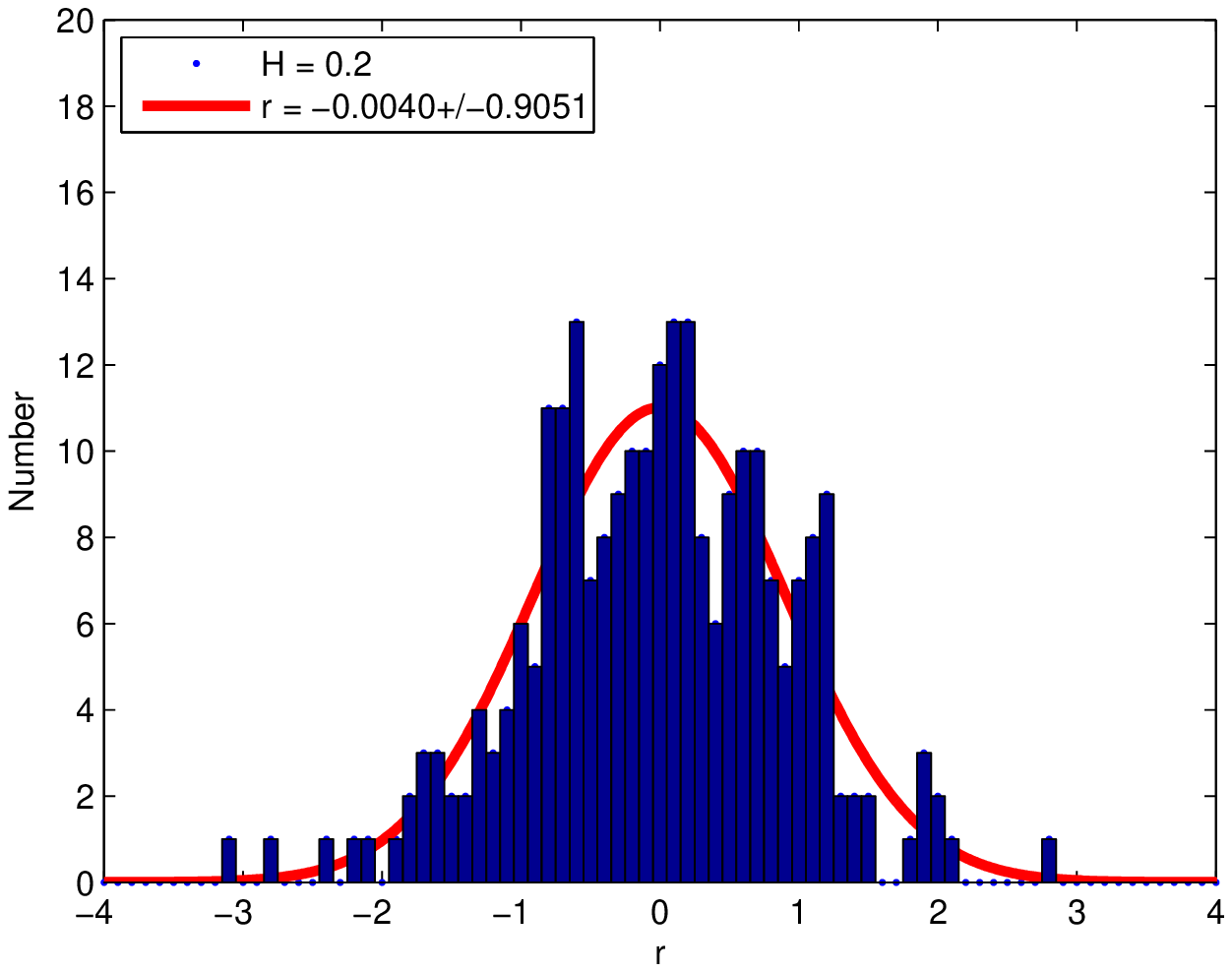}
\includegraphics[height=42mm,angle=0]{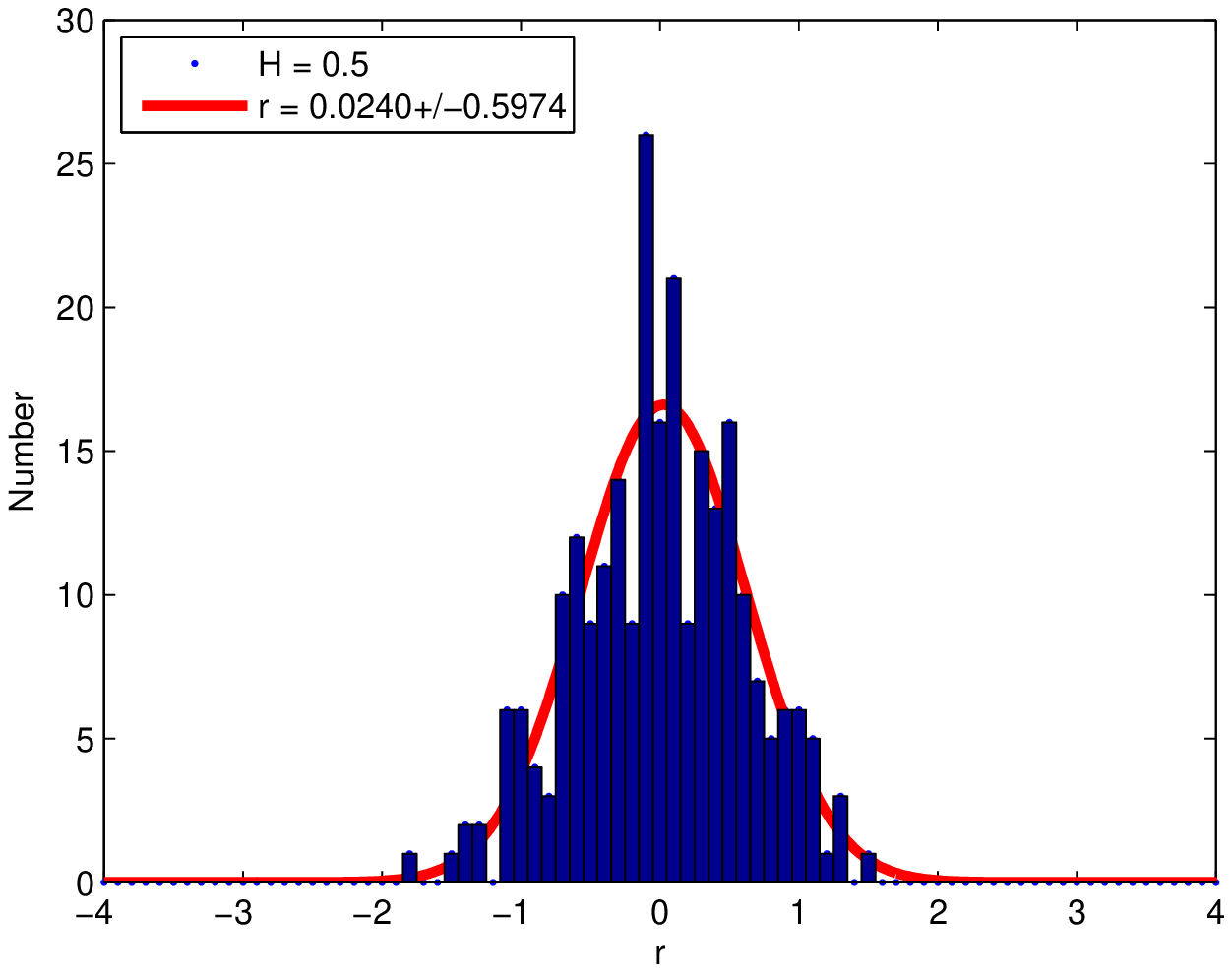}
\includegraphics[height=42mm,angle=0]{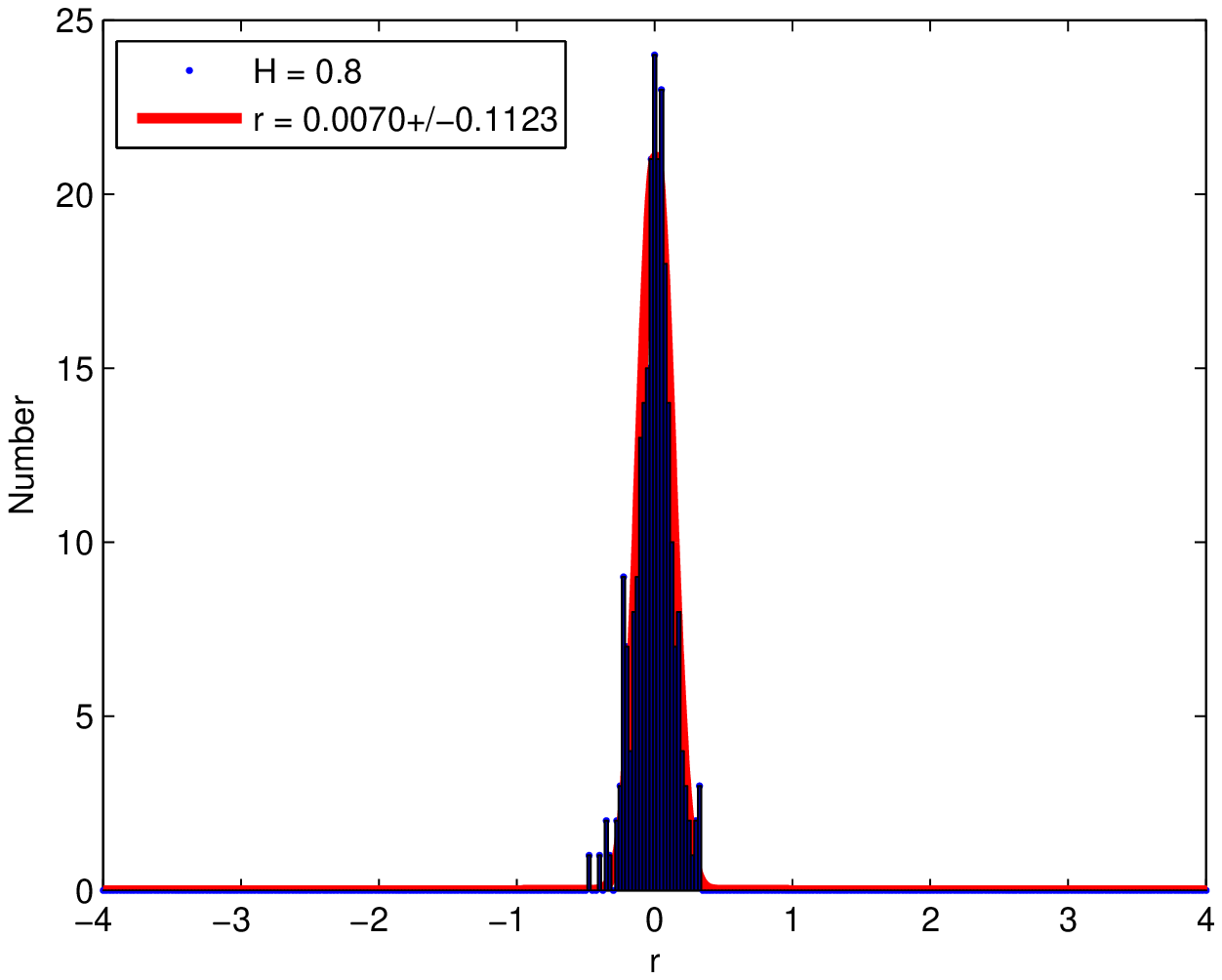}
\caption{Distribution of the normalized residual $r$ for $H=0.8, 0.5, 0.2$ shown in FIG. 1 and the Gaussian fitting is indicated by the red solid line with the fitting parameters also given in the figure.}
\label{b}
\end{figure*}

We first note that the normalized residuals $r=(\langle|\Delta F_{\Delta t}|\rangle-\hat{\langle|\Delta F_{\Delta t}|\rangle})/\sigma$, where $\hat{\langle|\Delta F_{\Delta t}|\rangle}$ corresponds to the best fit model, do not follow the standard Gaussian distribution as expected. Figure \ref{b} shows the distribution of these residuals and the corresponding fit with a Gaussian model. The width of the Gaussian fit is less than 1 for the three values of $H$ studied here. The expectation values of $S$ and $I$ are $H$ and $\textrm{log}(2/\pi)/2$($\simeq -0.0981$), respectively \cite{q14}. Although the best-fit values are close to these expectations, considering the small errors of these fitting parameters, the deviation of the best-fit values from their expectation appears to be significant.
%According to the works of Qiao, B. Q and Liu, S. M, (2013), we know that the excepted
%values of the slope and intercept of the fitting line are $S_0=H$ and $I_0=-0.0981$.
%In Figure \ref{a}, we can find that, for different $H$, the excepted values of the
%slope are not lied in between $S-S_{e}$ and $S+S_{e}$, but the excepted values of the
% intercept are in between $I-I_{e}$ and $I+I_{e}$. that sometimes the situation which the %expected values of the intercept exceed the interval from $I-I_{e}$ to $I+I_{e}$ can occur, the %case in Figure \ref{a} is only a special fitting result.
 % The results from the figure that
% the full width at half-maximum (FWHM) $\sigma_{\chi}$ of the $\chi^{,}$s Gaussian distribution %are not equal to unity except $H=0.2$. So the results
% for $H=0.8$ and $H=0.5$ are not consistent with the excepted result of theoretical prediction.

\begin{figure*}[htb]
\centering
\includegraphics[height=55mm,angle=0]{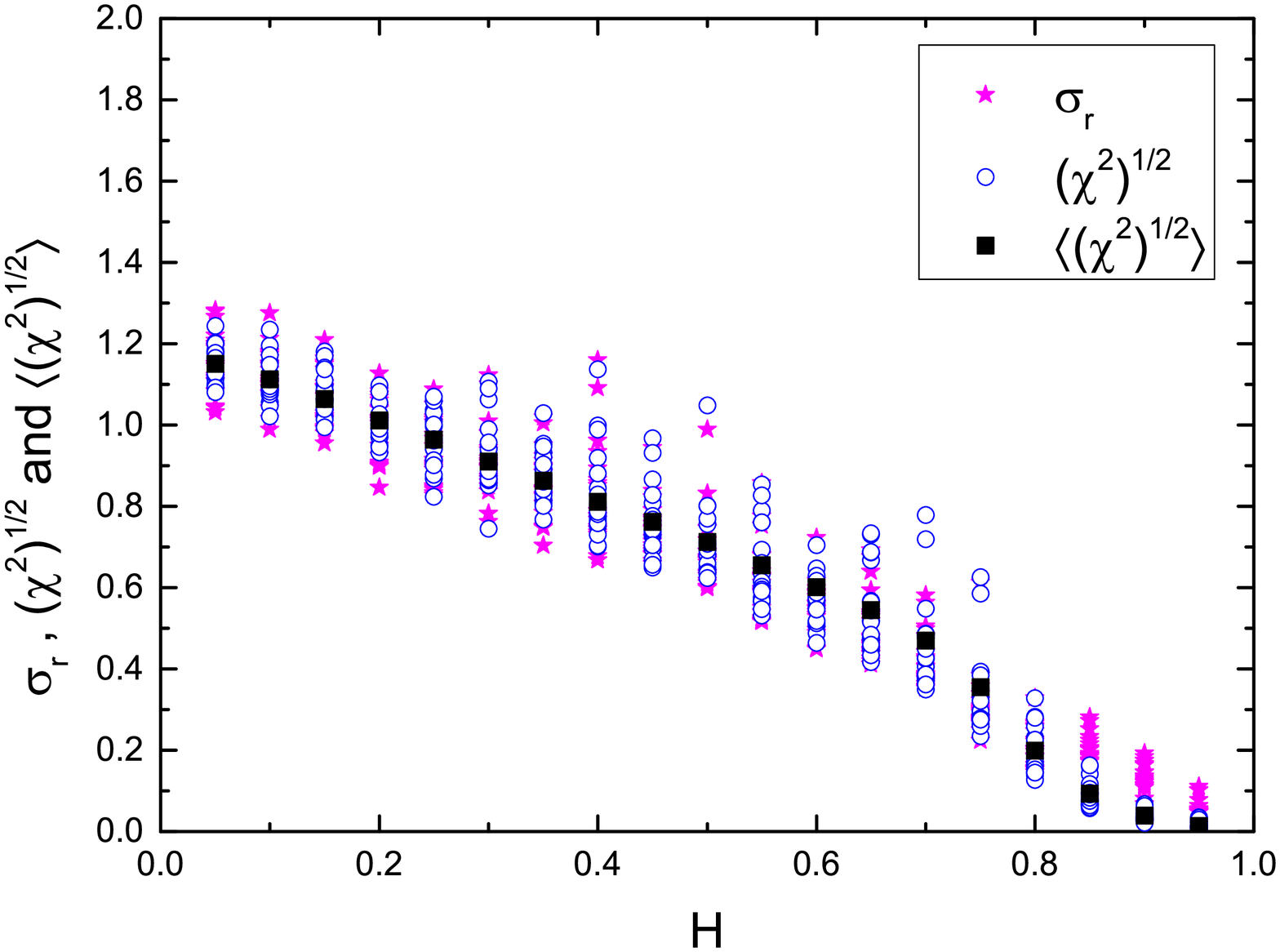}
\caption{The $H$ dependence of the square-root of the reduced $\chi^2$ and the Gaussian width of the normalized residual, $\sigma_r$, for 20 runs randomly selected from 2000 runs. The solid squares indicate the mean value of the square-root of the reduced $\chi^2$ for the 2000 runs.}
\label{f3}
\end{figure*}

\begin{figure*}[htb]
\centering
\includegraphics[height=55mm,angle=0]{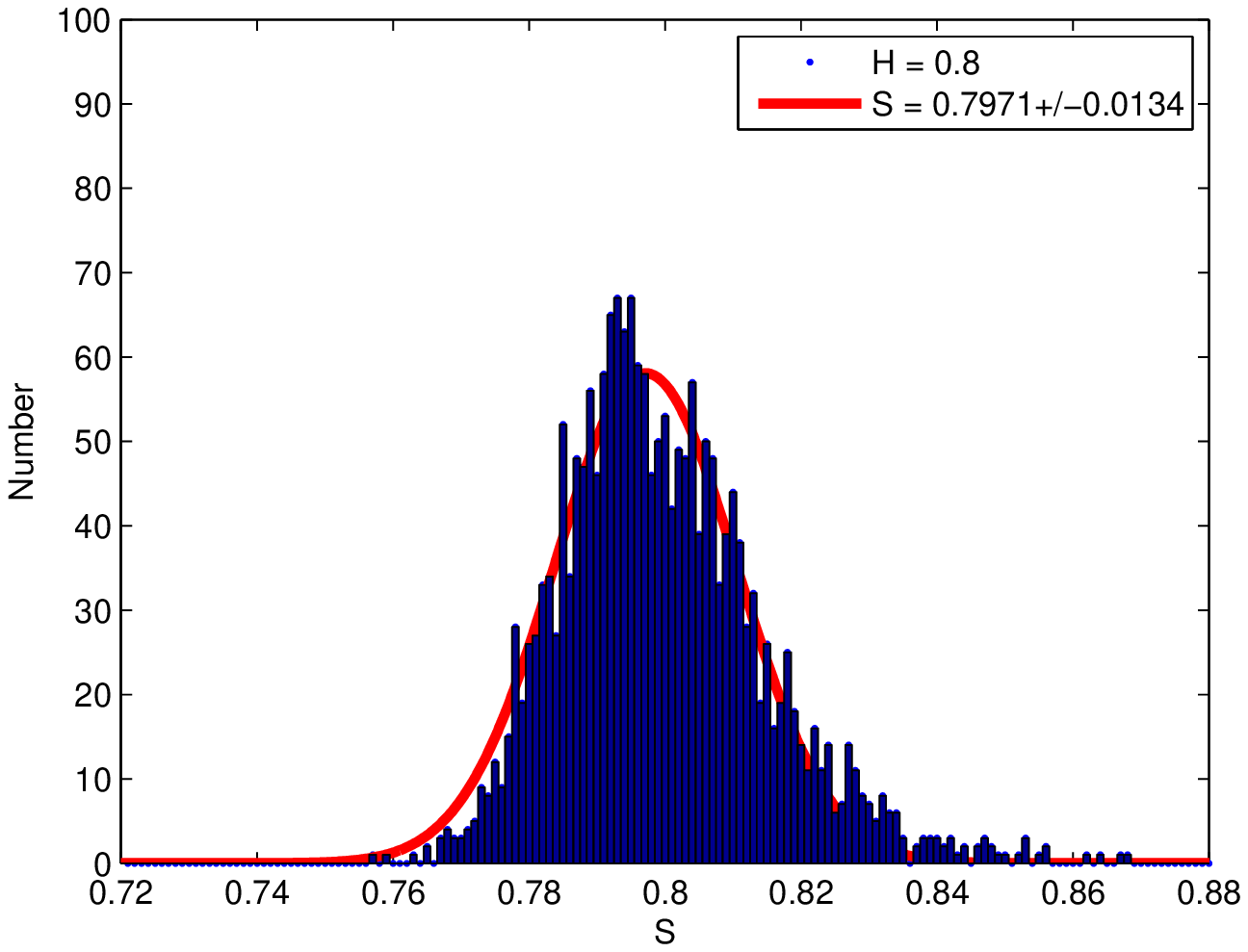}
\includegraphics[height=55mm,angle=0]{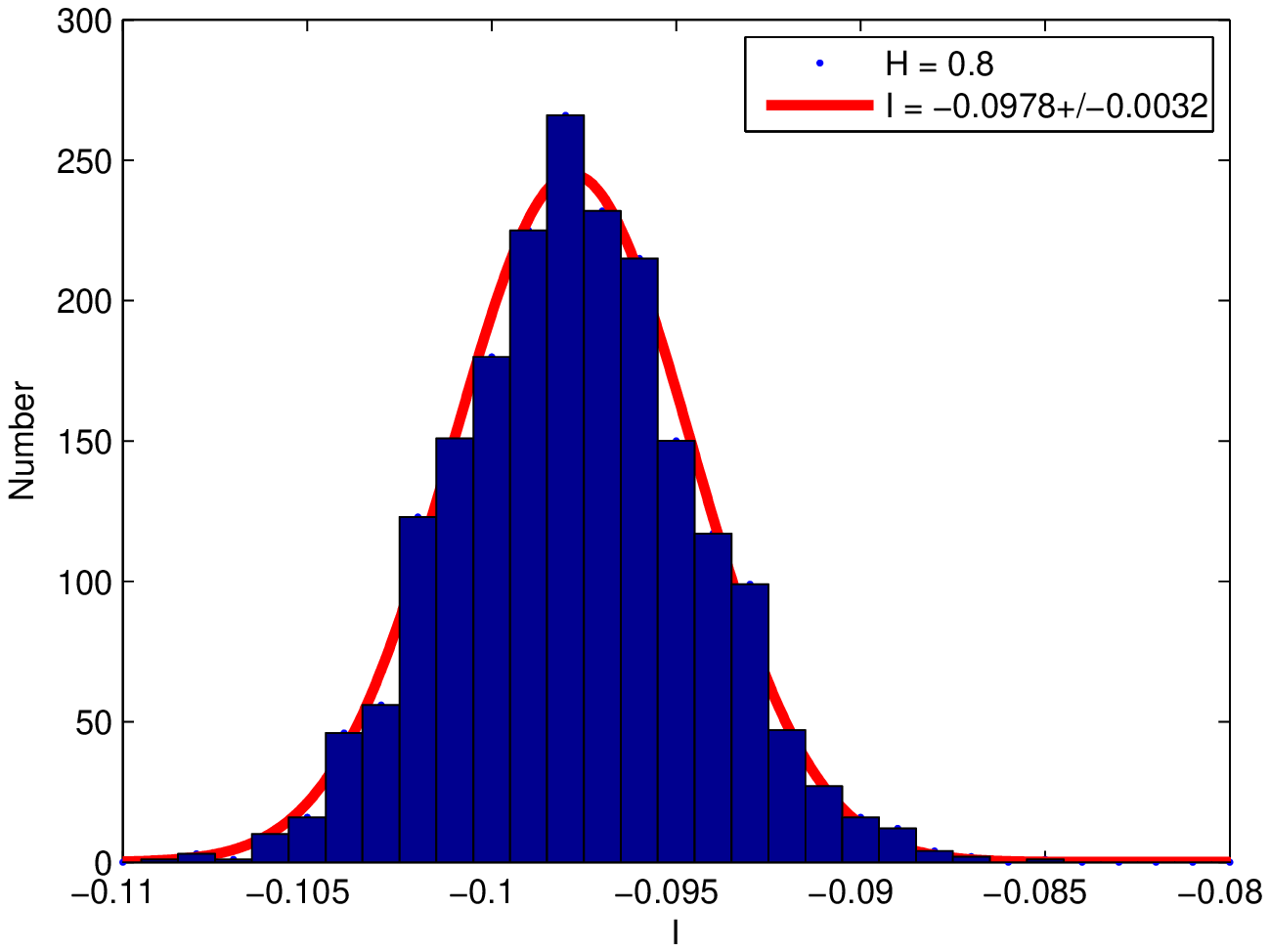}
\caption{Left: the distribution of the best-fit slope for 2000 runs with $H=0.8$ and the Gaussian fit.  Right: same as the left panel but for the intercept.
%the relation between standard deviation $\sigma_{\chi}$ of the Gaussian distribution of %normalized residual,  $\chi_{1}$ of the reduced $\chi$-squared root and the values of $H$. Each $H$ corresponds to 20 $\sigma_{\chi}$ or $\chi_{1}$ values.
}
\label{c}
\end{figure*}

To better understand these issues, for 19 different uniformly sampled values of $H$ between 0 and 1, the simulation and fitting above are repeated 2000 times.
Figure \ref{f3} shows the dependence of the square-root of the reduced $\chi^2$ on $H$ for 20 randomly selected simulations from the 2000 runs for each value of $H$. The stars show the corresponding widths of the Gaussian fits to the normalized residual, $\sigma_r$, as shown in FIG. 2, which are consistent with the value of the reduced $\chi^2$. We note that both the reduced $\chi^2$ and the width of the distribution of the normalized residual decreases with the increase of $H$. Figure \ref{c} show the distributions of the best-fit intercept and slope and their corresponding Gaussian fit for $H=0.8$.

\begin{figure*}[htb]
\centering
\includegraphics[height=55mm,angle=0]{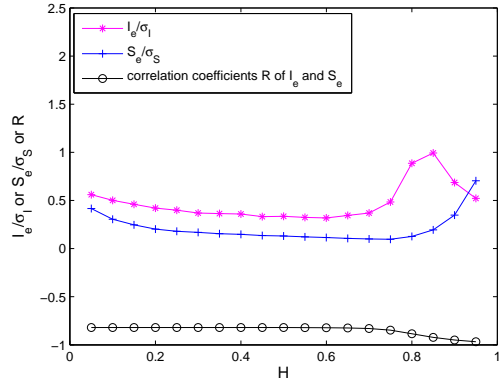}
\caption{%the relation between standard deviation $\sigma_{\chi}$ of the Gaussian distribution of %normalized residual,  $\chi_{1}$ of the reduced $\chi$-squared root and the values of $H$. Each $H$ corresponds to 20 $\sigma_{\chi}$ or $\chi_{1}$ values.
Dependence of the ${I_{e}}/{\sigma_{I}}$, ${S_{e}}/{\sigma_{S}}$ and $R$ on $H$ for 2000 runs of each $H$.}
\label{f5}
\end{figure*}

The distributions of $S$ and $I$ do not necessarily follow Gaussian distributions, especially for values of $H$ close to 0 or 1. In the following, we use the standard deviation of $S$ (denoted with $\sigma_S$) and $I$ ($\sigma_I$) to characterize the widths of these distributions.
Figure \ref{f5} shows the dependence of ${I_e}/{\sigma_I}$ and ${S_e}/{\sigma_S}$ on $H$.
The widths $\sigma_S$ and $\sigma_I$ are usually greater their errors $S_e$ and $I_e$ shown in FIG. 1, respectively. We note that $S_e$ and $I_e$ are determined by the standard deviation $\sigma$ and therefore do not change for the 2000 simulations and fittings with a given value of $H$\cite{p07}.
Figure \ref{f5} also shows the correlation coefficient R between $I_e$ and $S_e$, which corresponds to the non-diagonal term of the covariance matrix of the fitting parameters $S$ and $I$.
The fact that $R$ is close to $-1$ implies that the two fitting parameters are not independent. The anomalous behavior of $I_e/\sigma_I$ and $S_e/\sigma_S$ for $H>0.7$ may be related to the increase of the correlation between $I_e$ and $S_e$ as shown with $R$.
% Note that the values of the standard deviation $\sigma_{\chi}$ decrease with the increase of $H$ %and $\sigma_{\chi}$ and $\chi_{1}$ are approximately equal just as the theoretical expectation. %However, the values of $\frac{I_e}{\sigma_I}$ and $\frac{S_e}{\sigma_S}$ are
% not our excepted value that the specific value should be considered as floating around
% 1 for each $H$. We can also see that $R$ values show the strong correlation existed
% between $I_e$ and $S_e$. Those results are showed on the right bottom panel in Figure \ref{c}.

 % $\frac{S_e}{\sigma_S}$ that is a ratio of the slope's
 % error of the fits to standard deviation of the Gaussian distribution of
 % slope sets and $\frac{I_e}{\sigma_I}$ that is the ratio of the intercept's
 % error of the
 % fits to standard deviation of the Gaussian distribution of intercept sets.
%\begin{figure*}[htb]
%\centering
%\includegraphics[height=42mm,angle=0]{fig2a.eps}
%\includegraphics[height=42mm,angle=0]{fig2b.eps}
%\includegraphics[height=42mm,angle=0]{fig2c.eps}
%\caption{The histograms of the distribution of $\textrm{log}(\langle |\Delta F_t|\rangle)^{,}$s normalized residual $\chi$ with $H=0.8, 0.5, 0.2$ are indicated by blue bars and the Gaussian fitting are described by red solid lines. Their mean values and standard deviation values are also annotated on the figure.}
%\label{b}
%\end{figure*}

\begin{figure*}[htb]
\centering
\includegraphics[height=35mm,angle=0]{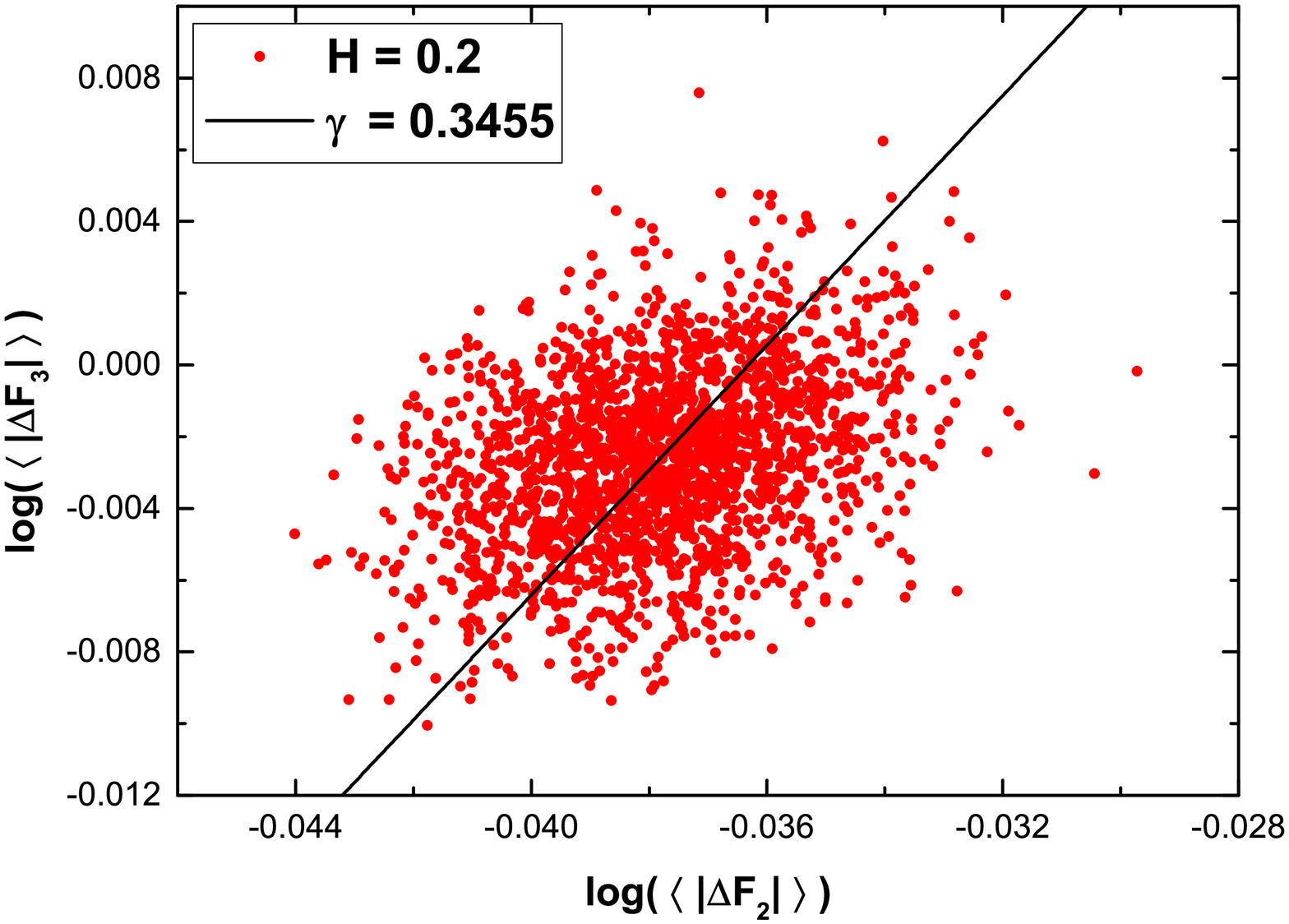}
\includegraphics[height=35mm,angle=0]{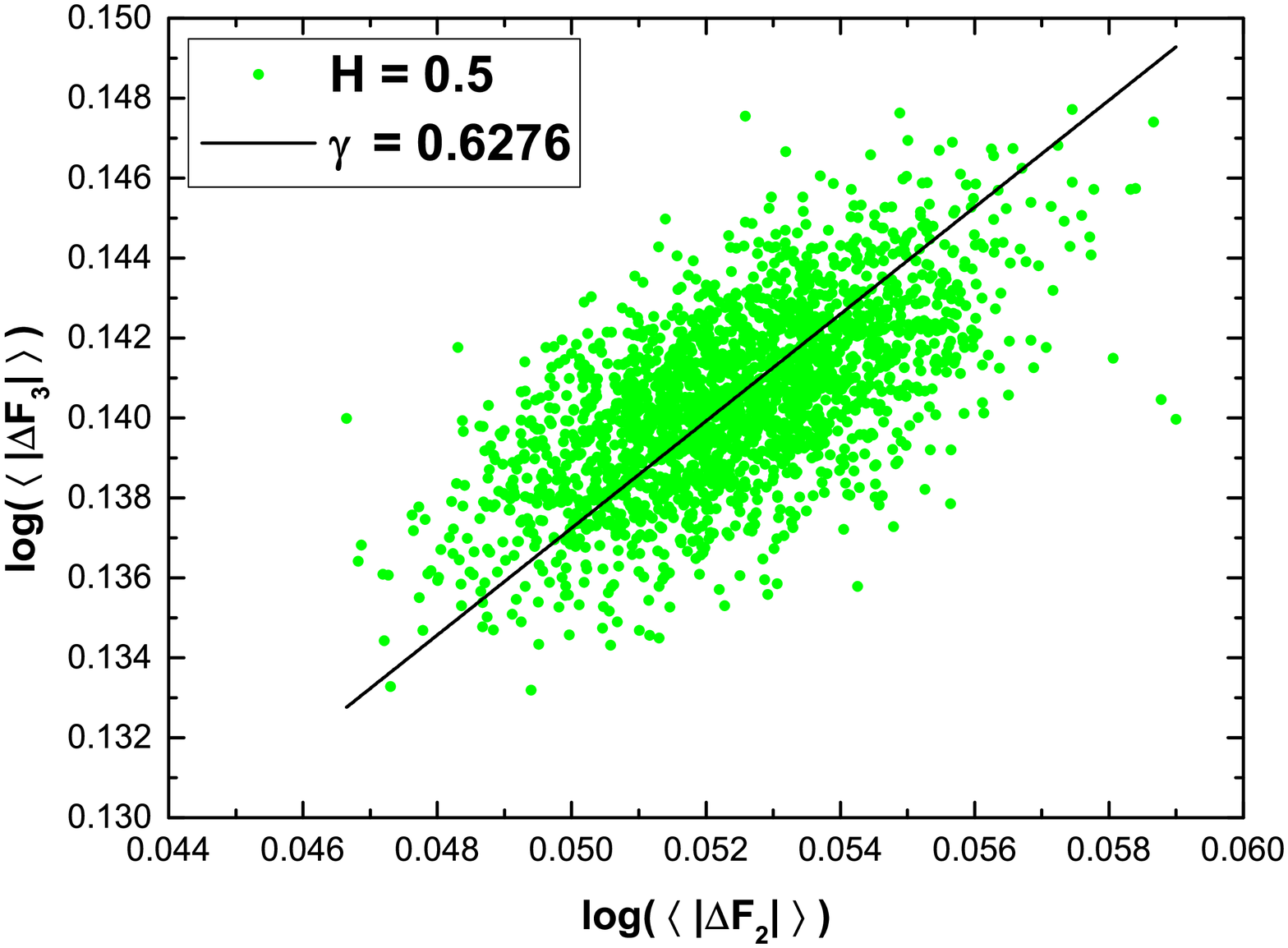}
\includegraphics[height=35mm,angle=0]{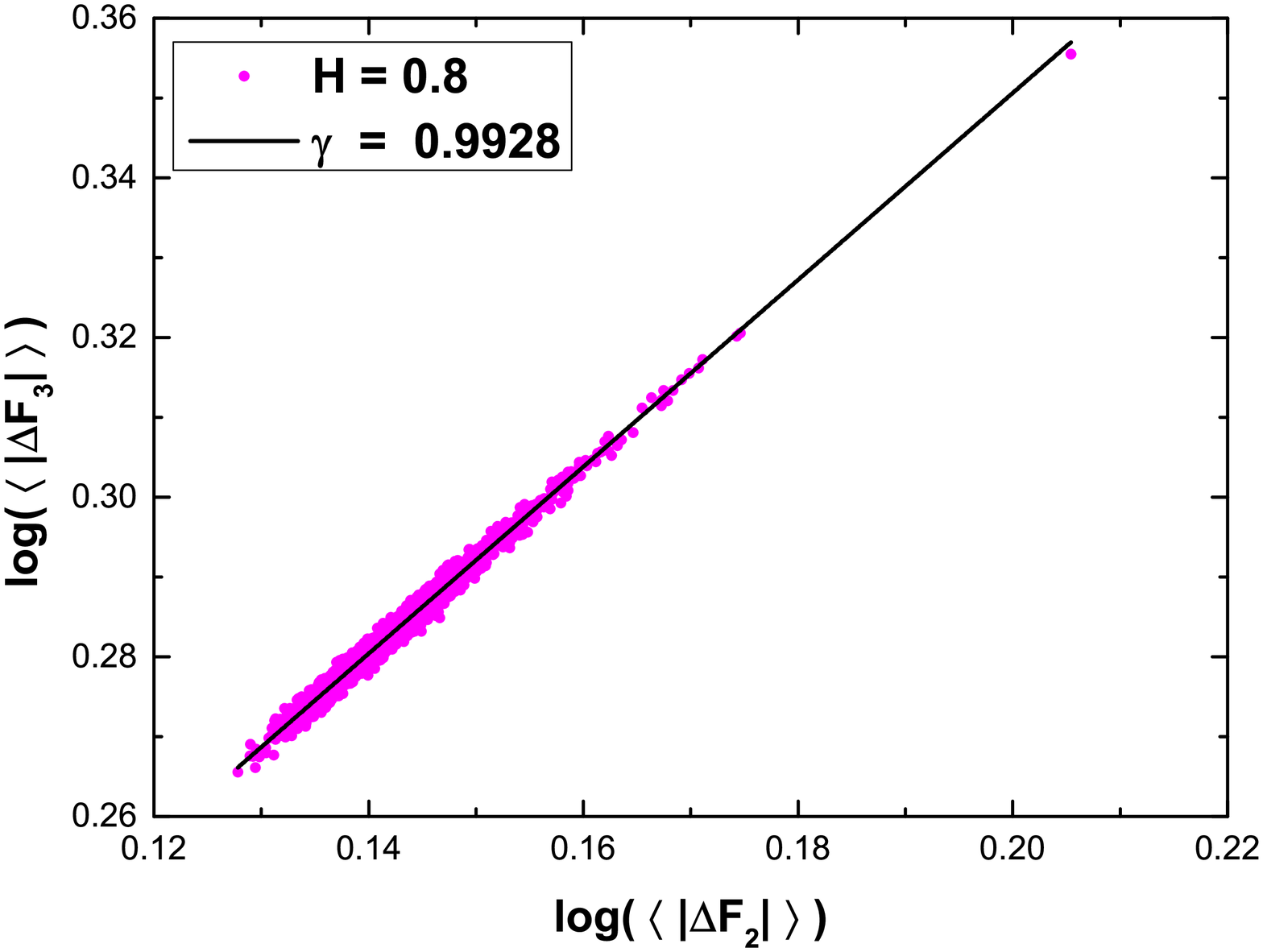}\\
\includegraphics[height=55mm,angle=0]{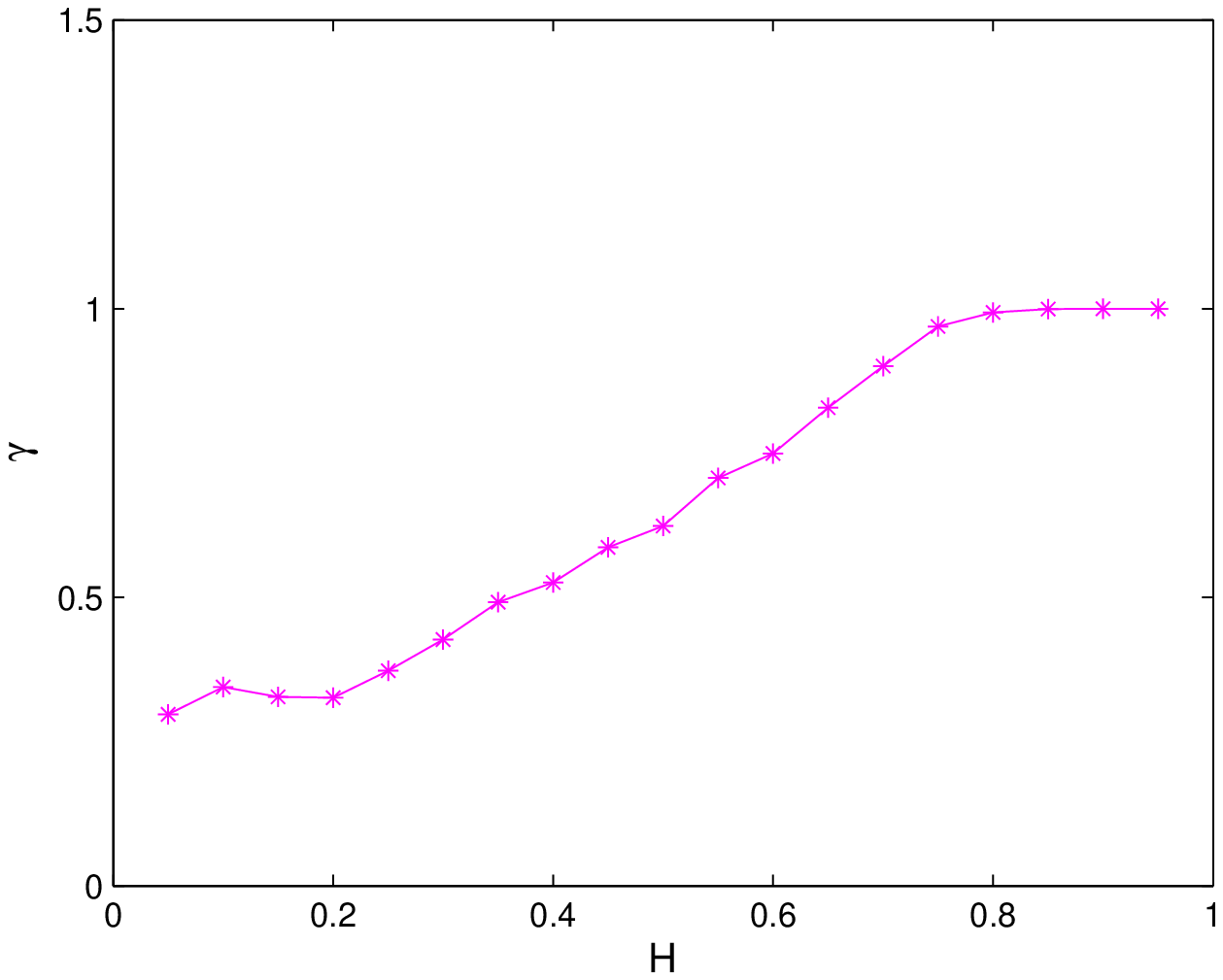}
\caption{Upper: correlation between log$(\langle|F(t+2)-F(t)|\rangle)$ and log$(\langle|F(t+3)-F(t)|\rangle)$ for 2000 runs of fBms with $H=0.8, 0.5, 0.2$. The corresponding correlation coefficients are indicated on the figure. Lower: dependence of the correlation coefficient $\gamma$ in the upper panel %between log$(\langle|F(t+2)-F(t)|\rangle)$ and log$(\langle|F(t+3)-F(t)|\rangle)$
on $H$.
%, and the value of $\gamma$ are calculated five times for each $H$.
}
\label{d}
\end{figure*}

The above results clearly show inconsistency in the evaluation of $H$ with the LS method applying to the $\textrm{log}(\langle|F(t+\Delta t)-F(t)|\rangle)$ vs $\textrm{log}(\Delta t)$ plot.
The strong correlation between the errors of the fitting parameters suggests that correlations among $\textrm{log}(\langle|F(t+\Delta t)-F(t)|\rangle)$ may play a role. For the simple LS method to be applicable, each rescaled range $\langle|\Delta F_{\Delta t}|\rangle$ needs to be independent of the others, which does not appear to be the case in our analysis since the rescaled ranges are derived from a given fBms. The upper panels of Figure \ref{d} show the correlation between log$(\langle|\Delta F_2|\rangle)$ and log$(\langle|\Delta F_3|\rangle)$ for several values of $H$. The lower panel shows the dependence of the correlation coefficient $\gamma$ on $H$. The correlation increases with the increase of $H$, reminiscent of the dependence of the reduced $\chi^2$ of the fitting on $H$ shown in Figure \ref{f3}.

\section{Results for another sampling method}

%To avoid our results originate from the sampling method for a given time series $F(t)$,
% in the following, we do the same work with the previous section by employing the other sampling %approach. As follows:
%\begin{equation}
%\langle |\Delta F_t|\rangle = ~\left\{
%             \begin{tabular}{ll}
%             $\displaystyle\sum_{i=0}^{N-1} |F(\Delta t + i*\Delta t)-F(i*\Delta t)|\over N$  ~~~~~   & $0 < H \leq 0.5$ \\
%             $\displaystyle\sum_{i=0}^{N^{\beta}-1} |F(\Delta t + i*\Delta t)-F(i*\Delta t)|\over N^{\beta}$  ~~~~~   & $0.5 < H < 1$
%             \end{tabular}
%             \right. ,
%\end{equation}
Qiao and Liu (2013) showed that the standard deviation of the rescale range has relatively simple expression for the following sampling method (case 3 of \cite{q14}):
\begin{equation}
\langle |\Delta F_{\Delta t}|\rangle = \frac{\displaystyle\sum_{i=0}^{N-1} [|F(\Delta t + i*\Delta t)-F(i*\Delta t)|]}{N}.
\label{s2}
\end{equation}
%This sampling method is the case 3 of Qiao, B. Q and Liu, S. M, (2013), and the errors of $y-$variable are also ensured in Qiao$\&$Liu$'$s (2013) paper. But, carefully,
The standard deviation of log$(\langle|\Delta
F_{\Delta t}|\rangle)$ is given by
%here differs from the equations given in Section 2 is obtained by other two expressions:
$$\sigma = N^{-1/2} {\rm log} (e) (\pi/2-1)^{1/2}
{\rm \ \ \  for\ \ \ } 0<H\leq0.5,$$
$$ \sigma = N^{-\beta/2} {\rm log} (e) (\pi/2-1)^{1/2} {\rm \ \ \ for\ \ \ } 0.5<H<1.0,$$ %Spontaneously,
which are larger than those for the sampling method in Section 2.
Figures 7-9 show the corresponding results, which are very similar to those in Section 2.
The most notable difference appears in $R$ and $\gamma$. With the current sampling method, the (anti-)correlation between the errors of the fitting parameters $S_e$ and $I_e$ is even stronger, while the correlation between log$(\langle|\Delta F_2|\rangle)$ and log$(\langle|\Delta F_3|\rangle)$ decreases slightly.
%Figure 6 and Figure 7 can be obtained by using the same way like before.
%Comparing with the results in Section 2, the similar results are obtained. It follows that both %our analysis suggest
%that our results do not result from the different sampling method and further confirm the %existent relevance between our sampling data.

\begin{figure*}[htb]
\centering
\includegraphics[height=80mm,angle=0]{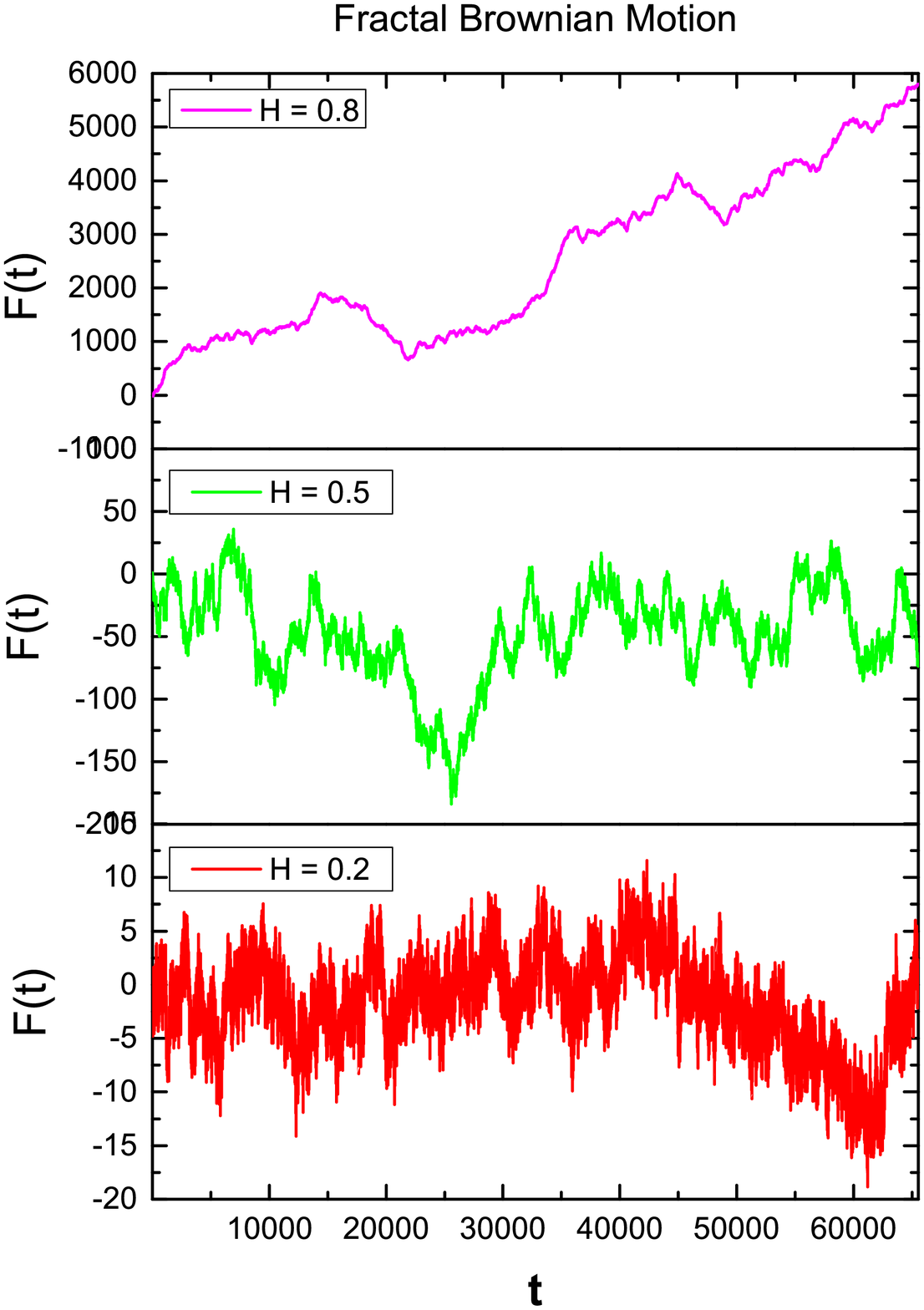}
\includegraphics[height=80mm,angle=0]{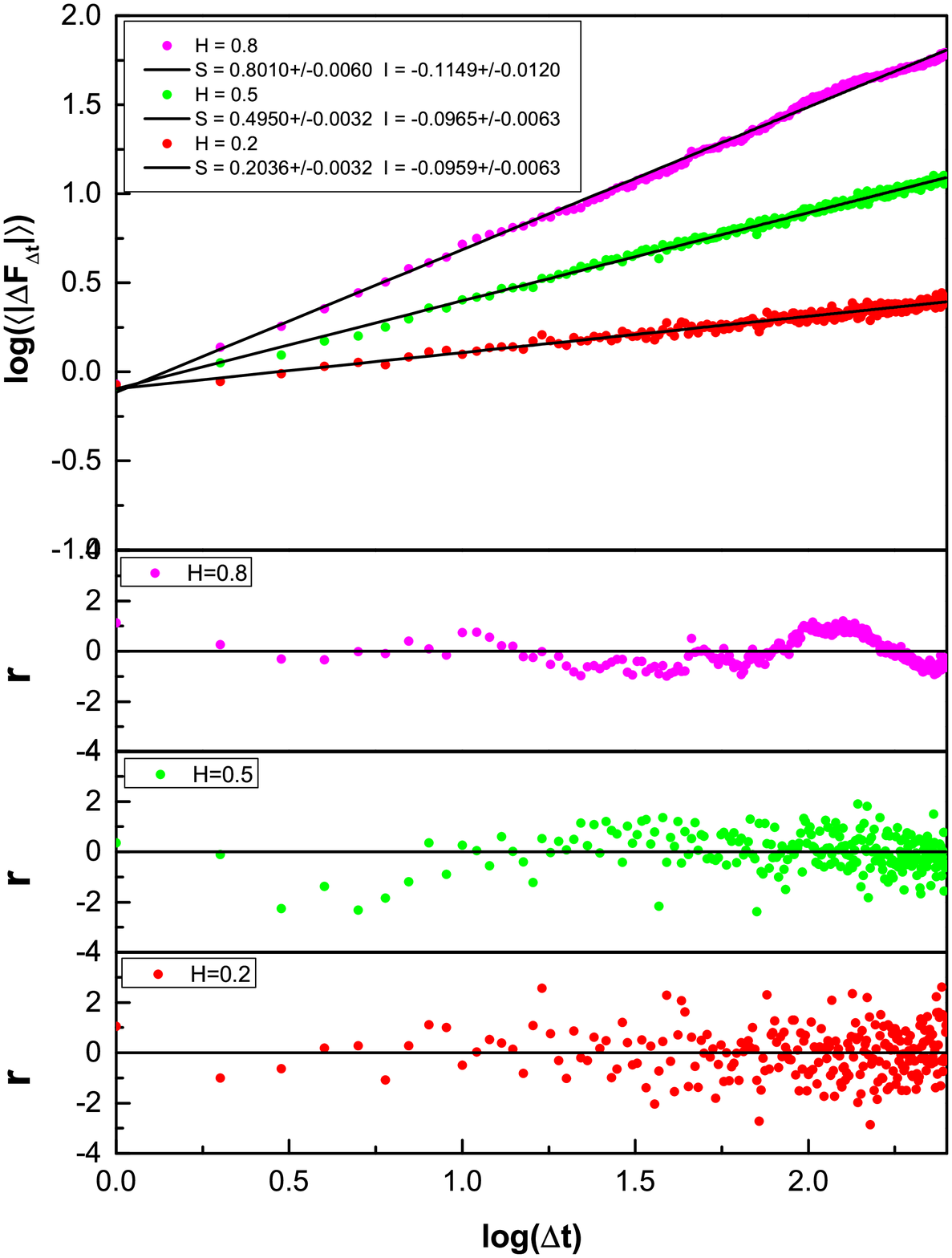}
\caption{Same as FIG. \ref{a} but for the sampling method given by equation \ref{s2}.}
\label{aa}
\end{figure*}

\begin{figure*}[htb]
\centering
\includegraphics[height=55mm,angle=0]{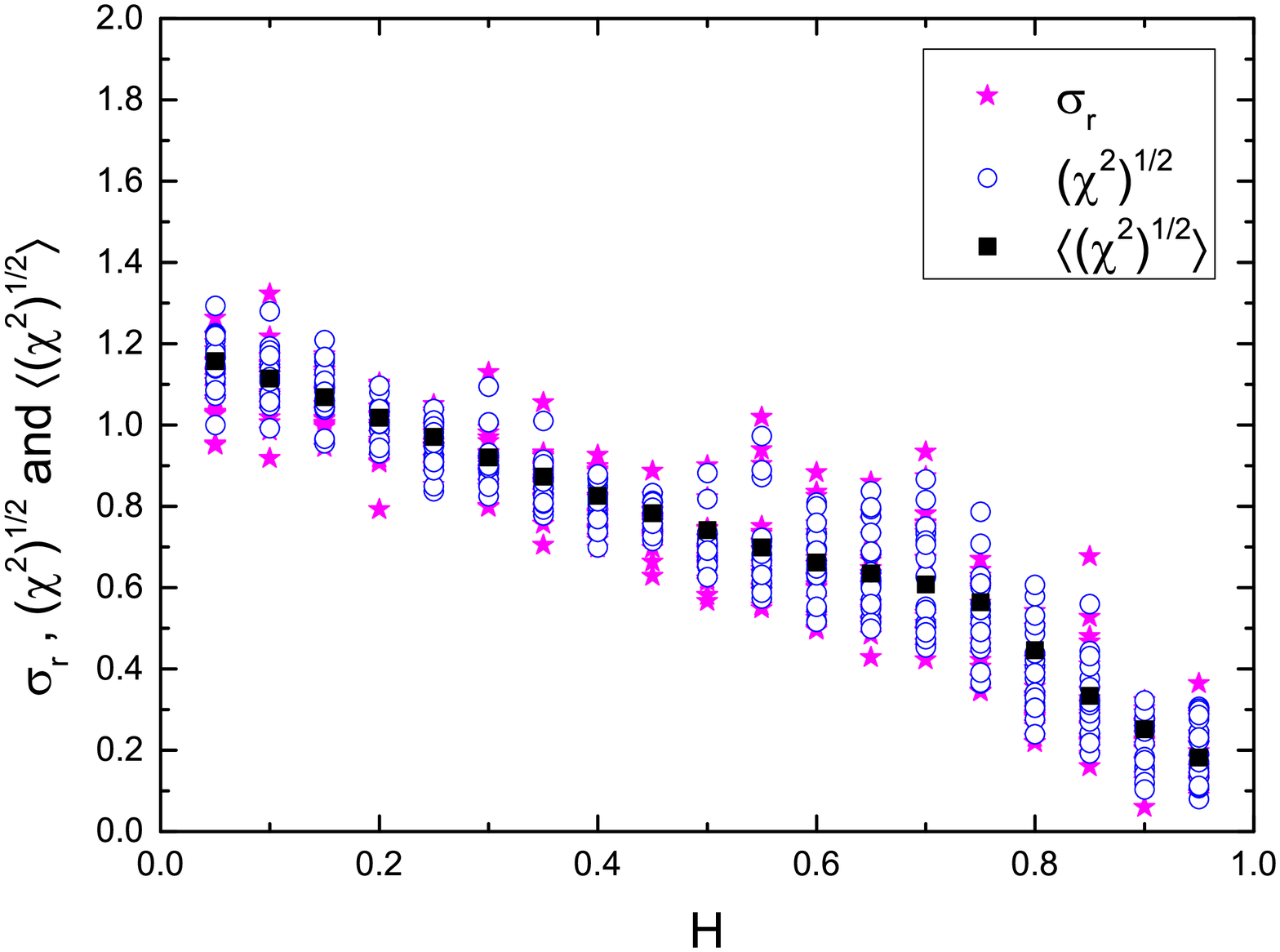}
\includegraphics[height=55mm,angle=0]{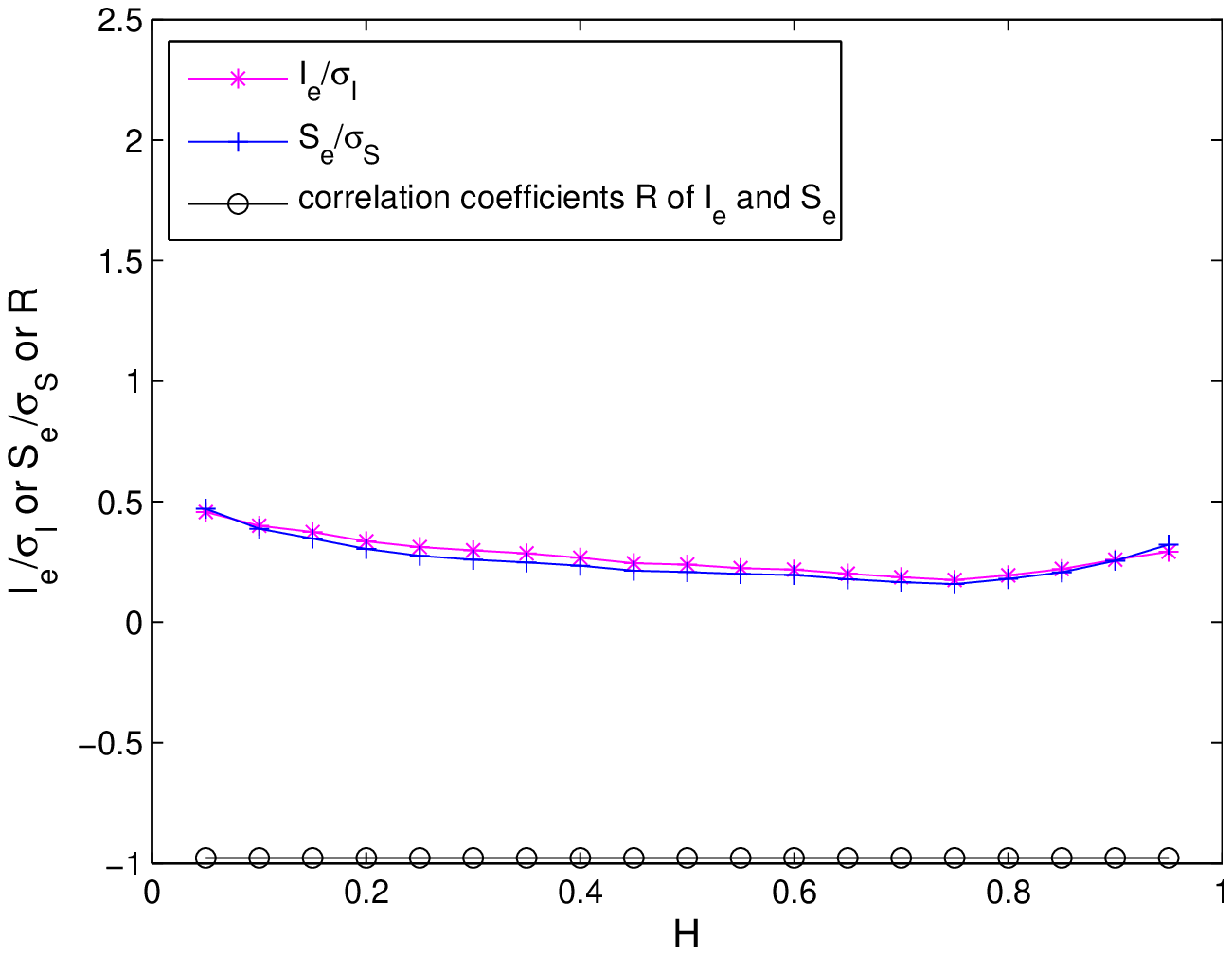}
\caption{Left: same as Fig \ref{f3} but for the sampling method given by equation \ref{s2}. Right: same as FIG. \ref{f5} for the sampling method given by equation \ref{s2}.}
\label{cc}
\end{figure*}

\begin{figure*}[htb]
\centering
\includegraphics[height=35mm,angle=0]{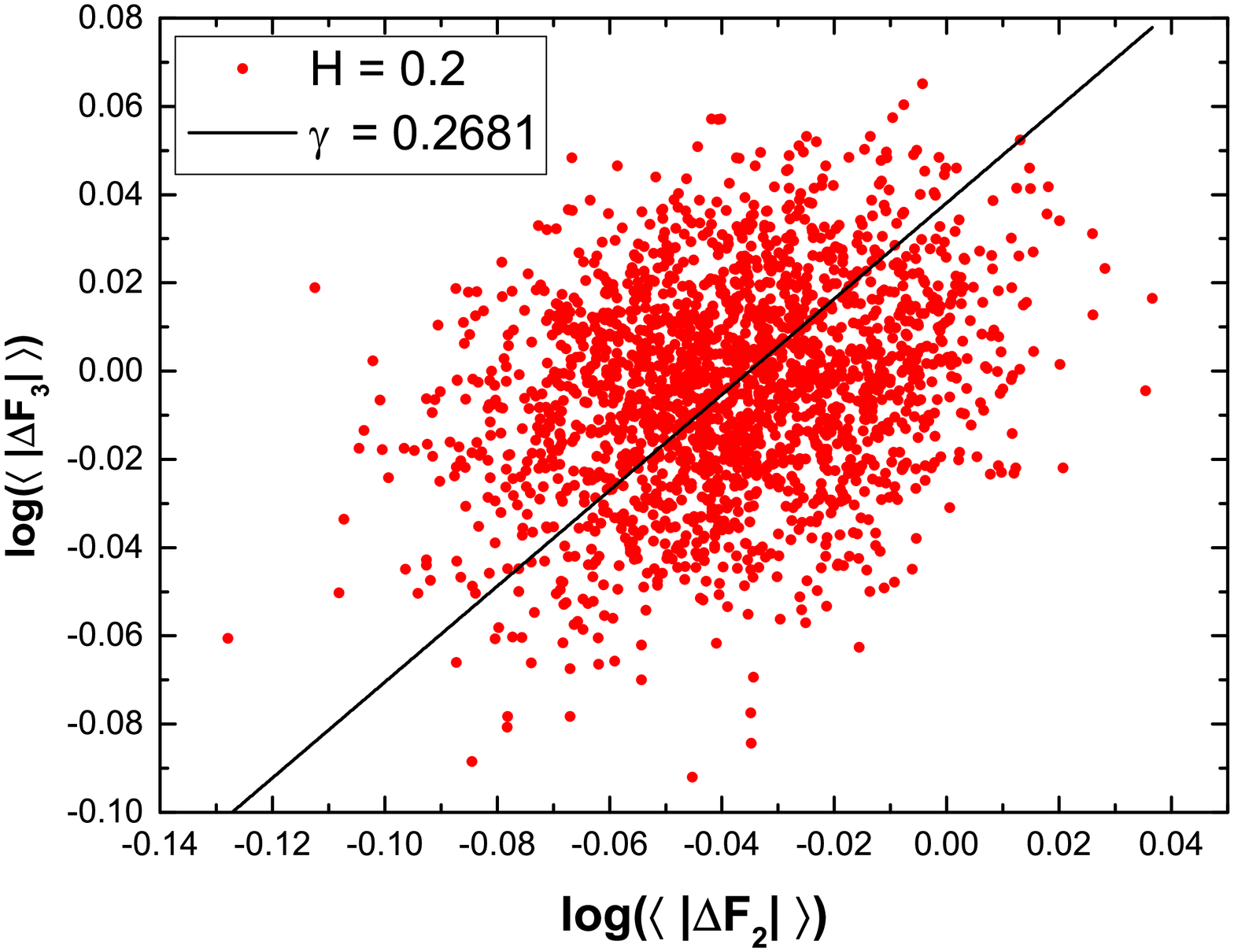}
\includegraphics[height=35mm,angle=0]{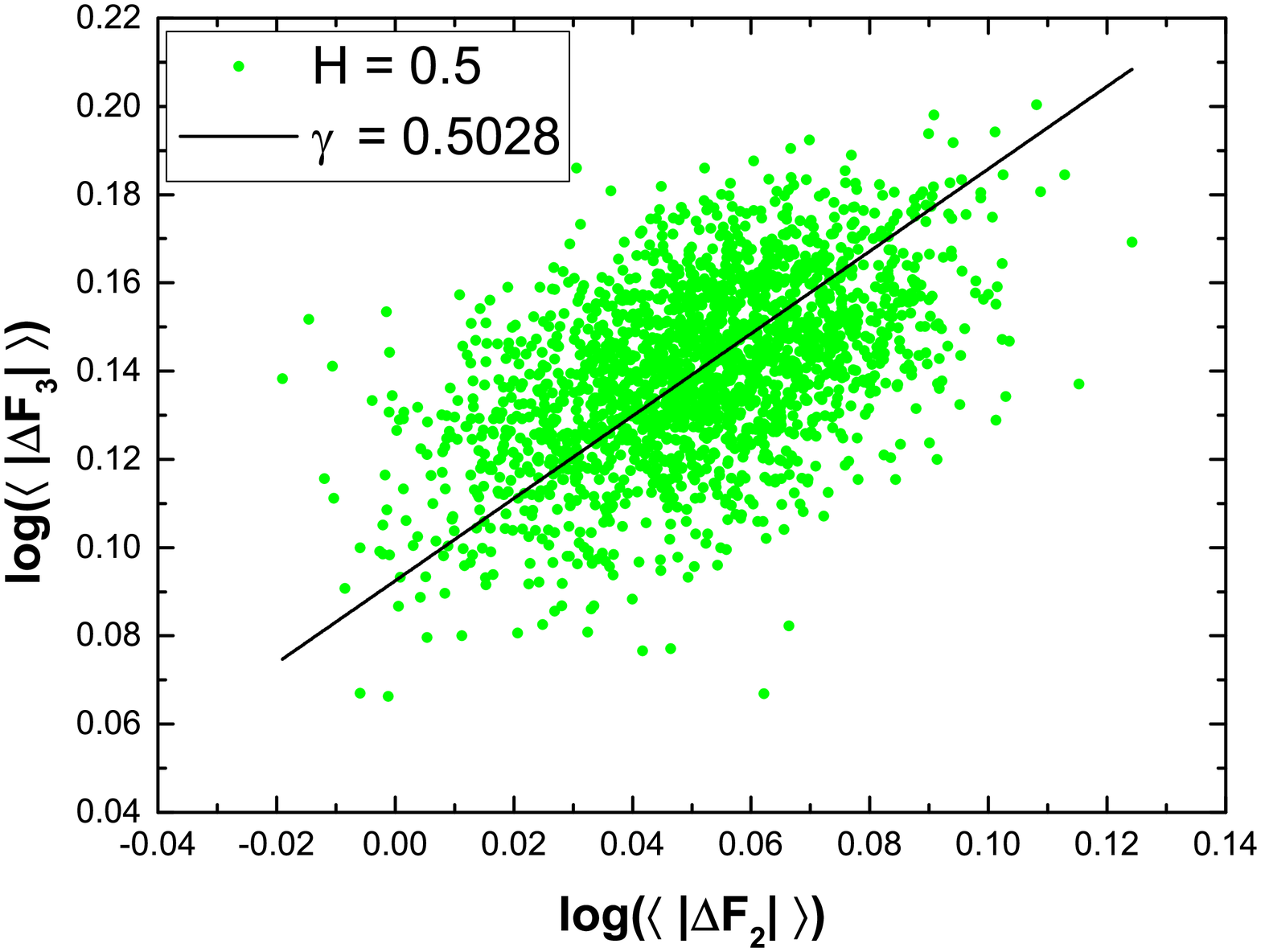}
\includegraphics[height=35mm,angle=0]{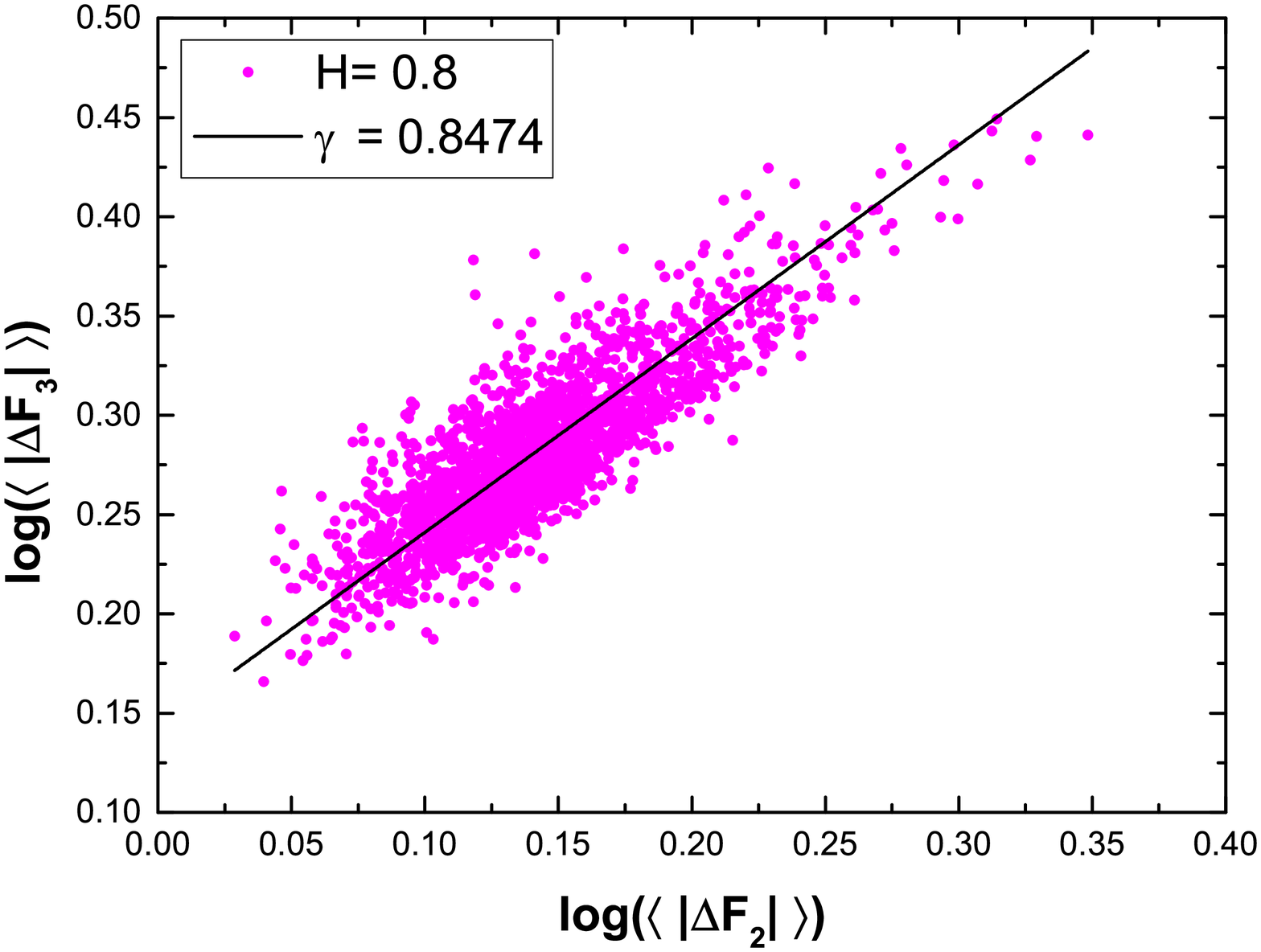}\\
\includegraphics[height=55mm,angle=0]{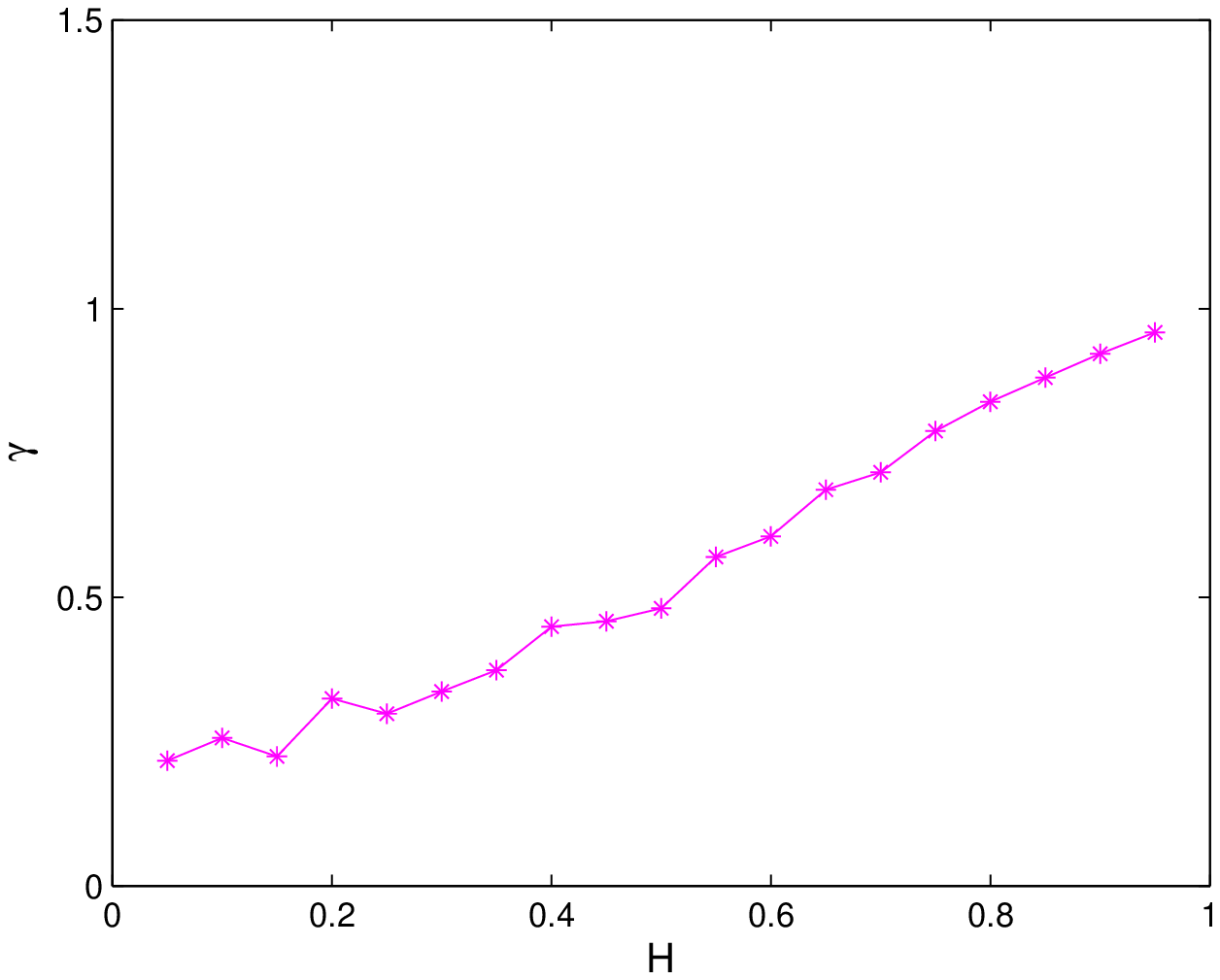}
\caption{Same as FIG. \ref{d} but for the sampling method given by equation \ref{s2}.}
\label{dd}
\end{figure*}

\section{Currency Exchange Rate between Euro and Dollar}

The results above show the limitation of the least square method in the evaluation of the dimension of fractal time series. However, we notice that the normalized residual of the LS fit shown in Figures. \ref{a} and \ref{aa} does not appear to depend on the time span, which also suggests the soundness of this procedure. One therefore may reinterpret the fitting parameters to obtain more self-consistent results. In this section, we use the currency exchange rate between euro and dollar in May and June of 2008 as an example to demonstrate how to obtain the dimension of fractal time series self-consistently.

\begin{figure*}[htb]
\centering
\includegraphics[height=54mm,angle=0]{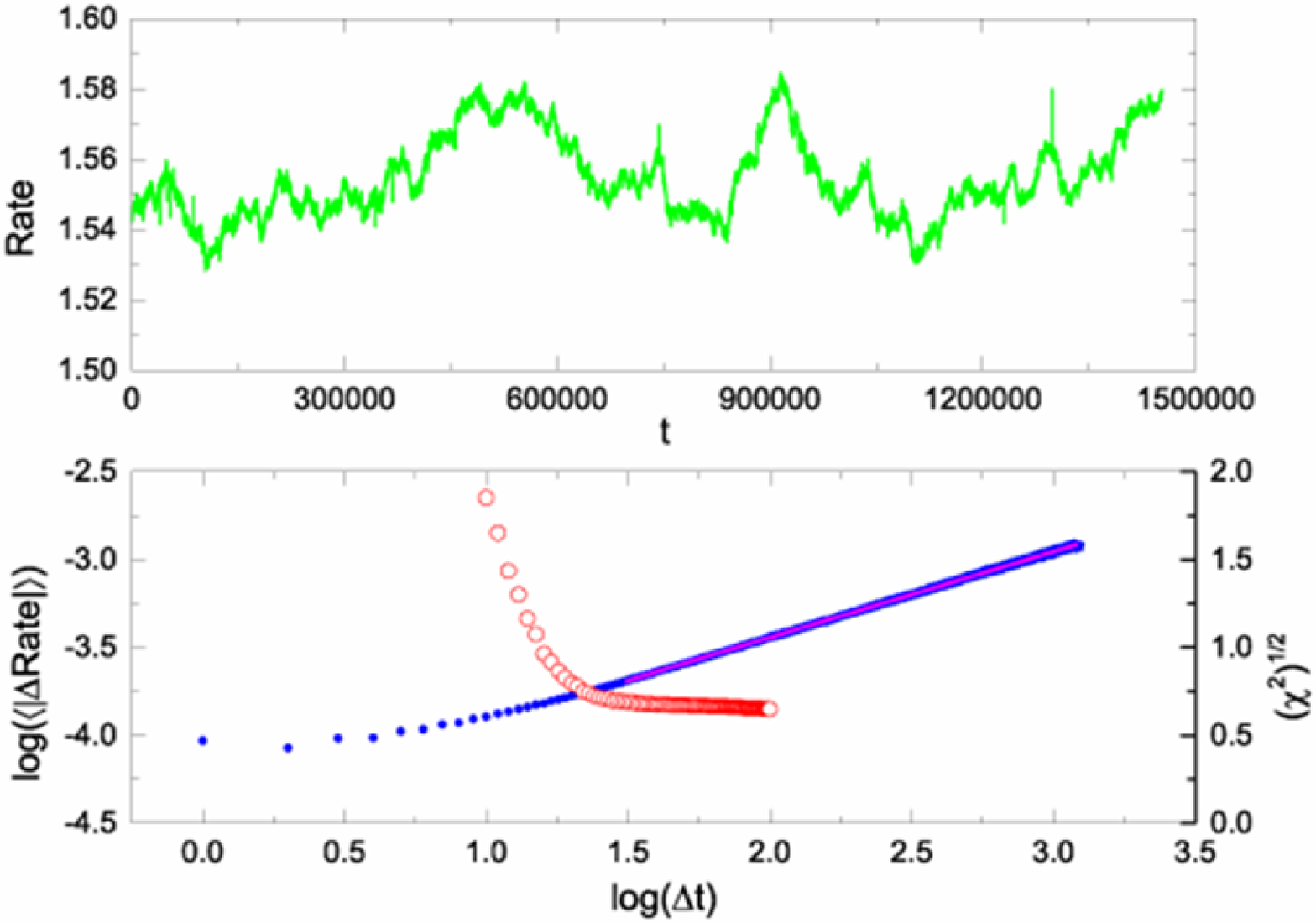}
\includegraphics[height=58mm,angle=0]{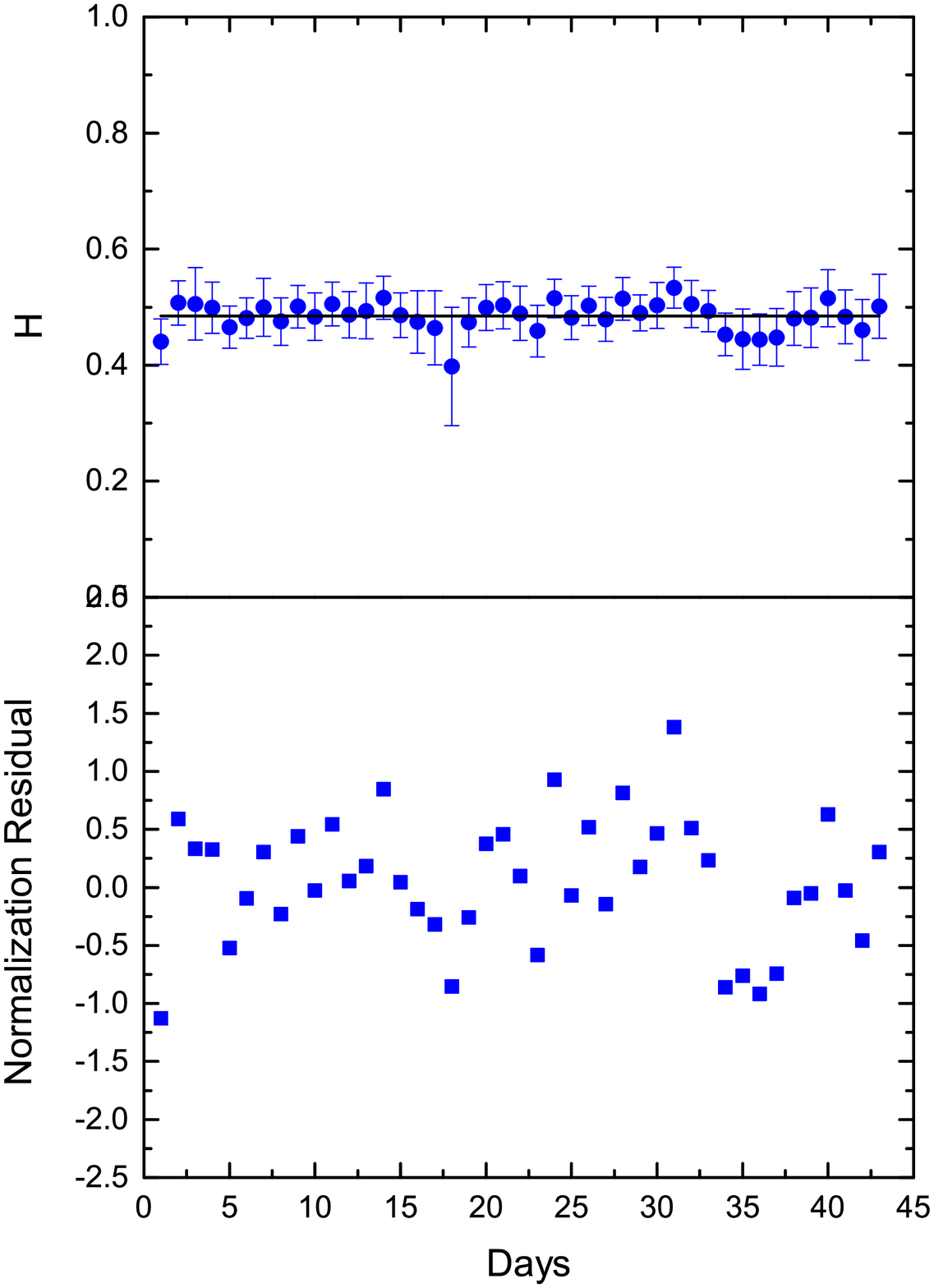}
\caption{
{\it Left:} Exchange rate between Euro and Dollar between May 1 and June 30 and its Hurst index analysis. The red open circles show the dependence of the square root of the reduced $\chi^2$ (indicated by the scale on the right) on the lower bound of the fitting range. The pink line indicates the best fit result with a lower bound of 30; {\it Right:} The Hurst index of exchange rate between Euro and Dollar for 43 days with extensive transactions.
}
\label{f10}
\end{figure*}

The top-left panel of Figure \ref{f10} shows the exchange rate as a function of the number of transaction $t$. It covers a duration from May 1 to June 30. The market was closed during the weekends. In total we have 43 days with extensive transactions. Since the sampling method corresponds to the case 4 of Qiao, B.Q and Liu, S.M, (2013) has the lowest standard deviation for the rescaled range, we will only consider this case in the following analysis.

The bottom-left panel of Figure \ref{f10} shows the dependence of the rescaled exchange rate range on the span of the transactions $\Delta t$. Because the error bars are very small, they are almost invisible in the figure. It is evident that the rescaled exchange rate range has some curvature below a few tens of transactions. Within $\sim4$ transactions, the rescaled range does not appear to vary with the transaction span, implying complete randomness similar to white noises. However, above a transaction span of $\sim30$, the rescale range can be fitted linearly with a slope of $0.489\pm0.002$, corresponding to a fractal dimension of 1.511. The reduced $\chi^2 = 0.48$, which is consistent with results shown in Figure \ref{f3}.

To further demonstrate the consistency of the above procedure, we analyze the data for the 43 days separately. The top-right panel of Figure \ref{f10} shows Hurst indexes, where the errors have been corrected using results in Figure \ref{f5}. The bottom-right panel of Figure \ref{f10} shows the normalized residual, which is consistent with a standard Gaussian distribution.

\section{Conclusion and Discussion}

The scaling of the rescaled range $\langle|\Delta F_{\tau}|\rangle$ with the time span $\tau$ plays an important role in quantitative analysis of fractal time series $F(t)$. The Hurst exponent is defined as the exponent of a power-law scaling. However, the consistency of methods of the Hurst exponent measurement has not been explored. One of the commonly used method is to apply the lease square method to the log$(\langle|\Delta F_{\tau}|\rangle)$ vs log$(\tau)$ plot. Using the mathematically well-defined fractal Brownian motions as a standard model, we investigate the limitation of the LS method.

It is shown that, due to the correlation among the rescale range for the sampling methods adopted in the paper, the LS method does not give self-consistent results. The two sampling methods have the most comprehensive usage of the available data and relatively simple expressions for the standard deviation of the rescale range. Other sampling methods are relatively less practical and less efficient. In principle, one may remedy the above procedure by taking into account the correlation among the rescale range into account in the fitting process. In that case, instead of fitting a simple function log$(\langle|\Delta F_{\tau}|\rangle)$ vs log$(\tau)$, one also needs to know the covariance matrix of log$(\langle|\Delta F_{\tau}|\rangle)$ in advance. These calculations are above the scope of the current investigation.

The fact that the normalized residual of the LS fit does not appear to depend on the time span also suggests the soundness of this procedure. The LS method may still be adopted in practical application by modifying the interpretation of the fitting result slightly with the results in this paper. First the reduced $\chi^2$ should have a best fit value given in figures \ref{f3} and \ref{cc} instead of 1 in the simplest LS method. Second the errors of the fitting parameters need to be corrected with the results shown in figures \ref{f5} and \ref{cc}.

\section*{Acknowledgements}
This work is partially supported by the NSFC grants 11173064, 11233001,
11233008 and 11063003.

{}

\end{document}